# Teaming up with information agents

Dr. Jurriaan van Diggelen
Dr. Wiard Jorritsma
Dr. Bob van der Vecht

jurriaan.vandiggelen@tno.nl, wiard.jorritsma@tno.nl, bob.vandervecht@tno.nl

TNO, the Netherlands

**Introduction**

Recent developments in Artificial Intelligence (AI) have led to impressive results in machine learning and pattern recognition, but have also led to the insight that AI hardly ever functions in isolation [1]. Most practical AI applications involve humans, e.g. for providing instructions, for correcting the machine if needed, for interpreting the machine's outcomes. A recent article [2] summarizes this as "*no AI is an island*", and argues that AI agents should be endowed with *teaming intelligence* that allows them to team up with humans.

Whereas teaming skills come naturally to humans, coding them into a computer has proven difficult. It involves (among others) making the computer decide which information to share with its teammates, which actions to undertake to complement those of its teammates, and how to explain its behavior to others that depend on it. Such team behaviors change over time, and depend on the context, competencies and performance of the involved actors, risks, and the state of others.

Despite the intricacies involved, we can observe patterns in team behavior which allow us to describe at a general level how AI systems are to collaborate with humans [3]. Whereas most work on human-machine teaming has focused on physical agents (e.g. robotic systems), our aim is to study how humans can collaborate with information agents. We propose some appropriate team design patterns, and test them using our Collaborative Intelligence Analysis (CIA) tool.

**Information agents for intelligence collection**

Information agents are special in the sense that their entire OODA (Observe-Orient-Decide-Act) cycle takes place in the information environment (cyberspace). Physical agents, on the other hand, observe and act in the physical environment which places physical boundaries on the execution speed which are not present in the information environment. Examples of information agents are cyberdefense systems (such as spam filters, and virus scanners), offensive cyber agents (such as malware, phishing software), and intelligence collection tools. We adopt the latter as a use case.

The intelligence collection process can be divided in the following phases: (1) *Direct,* in which potentially relevant information sources are determined (e.g. twitter feeds, camera streams); (2) *Collect,* in which the information sources are searched for relevant items (e.g. a particular tweet that is considered relevant); (3) *Process*, in which relevant items are checked for reliability and potentially classified as support for a certain hypothesis on the nature of the incident. This may lead to new information questions that form the start of a next cycle.

Intelligence analysis requires the processing of vast amounts of data from various sources and is loaded with ethical concerns, as collecting data often involves an intrusion of privacy and cannot be done without a good reason. This makes it a representative use case for human-agent teaming, involving ethical and legal aspects under time pressure and uncertainty.

**Team design patterns for information agents**

In order to define different human-agent team configurations and be able to transform from one form to the other, we have developed general team design patterns, which will be applied to the intelligence collection use case. Figure 1 presents some elementary ones to get the reader acquainted with the idea.

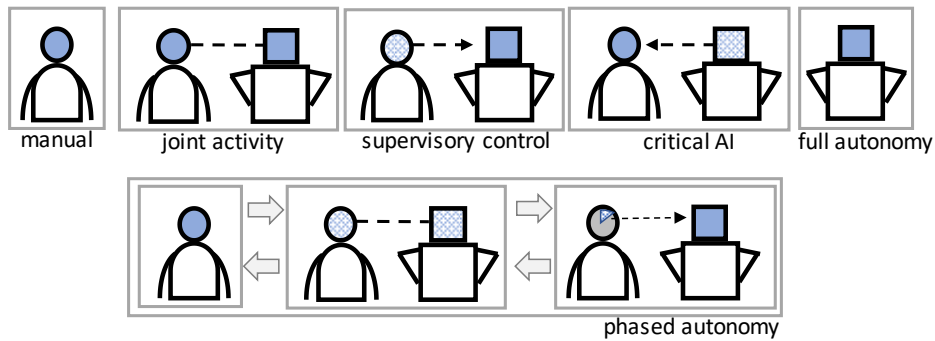

*Figure 1: Basic team design patterns for information agents*

The simple graphical teamwork pattern language introduced in [3] allows us to specify basic ways in which humans and/or systems perform work. A solid-filled head represents cognitive work carried out by the actor. A dotted-filled head represents indirect work, i.e. work that does not directly contribute to task-completion, such as in the *supervisory control* pattern where the human monitors the machine. More interesting patterns emerge when these basic patterns are nested in time, such as in the *phased autonomy* pattern, where the human does the task manually, but chooses to hand work over to the machine by going into a more autonomous mode. Noteworthy are: (1) the transition between manual and autonomous is not direct but intervened by a handover phase; (2) even in the "highly autonomous" pattern, a small part of the human's cognitive resources go to monitoring the machine to prepare for a possible handover to manual mode. These subtleties give important guidelines for designing the social behaviors into AI systems which are often overlooked.

**Collaborative Intelligence Analysis (CIA) tool**

In order to experiment with different forms of human-agent teaming in the intelligence domain, we have developed the CIA tool (Figure 2), which was built upon the MATRXS (Man Agent Teaming Rapid eXperimentation Software) framework currently under development at TNO. In the CIA tool, users are presented with a scenario (e.g., an unknown drone was spotted near a military terrain) and a number of hypotheses that possibly explain the scenario (e.g., preparations for an attack, espionage, false alarm). The goal is to find out which hypothesis is most likely by collecting and processing information from various information sources.

The tool can be run in different modes to experiment with different team design patterns. Three basic modes are currently supported: manual (in which the user performs the analysis manually), autonomous (in which (simulated) intelligent information agents collect and process information without human intervention), and collaborative mode (in which the user and the information agents perform the analysis together).

Performing the analysis in manual mode is a time-consuming and tedious task. The information agents can greatly speed up the process and potentially improve its accuracy by considering and cross-referencing more data than would be humanly possible. However, the autonomous mode reveals several downsides to fully autonomous intelligence analysis: (1) The human has no way of correcting any mistakes that the agents make; (2) The human has no control over which sources the agents access. Some sources are sensitive and may only be accessed under certain circumstances; (3) The human cannot use his or her experience, intuitions and general knowledge to guide the agents in a promising direction.

Therefore, we need a form of human-agent collaboration that leverages both the fast big data processing capabilities of the agents, and the domain and common-sense knowledge of the human. Our collaborative mode is a first step in this direction. We are currently designing more detailed team design patterns, and will implement and empirically evaluate them using our CIA tool in order to determine which patterns produce the most effective human-information agent teaming.

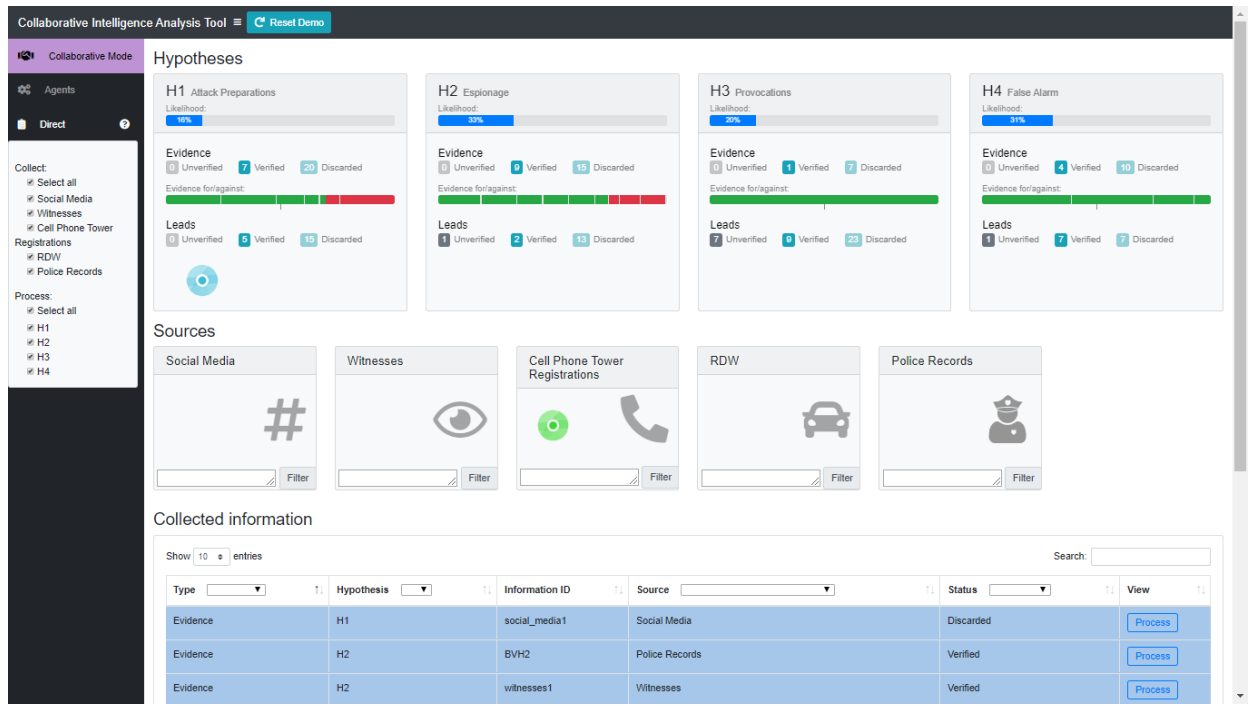

*Figure 2: Screenshot of the Collaborative Intelligence Analysis (CIA) tool*